# Impact of counteracting vehicles on the characteristics of a smart city transport system[1]

Nikita V. Bykov

*Russian University of Transport, Moscow, Obraztsova st., 9b9, Russia*
bykov@bmstu.ru

The development of smart city transport systems, including self-driving cars, leads to an increase in the threat of hostile interference in the processes of vehicle control. This interference may disrupt the normal functioning of the transport system, and, if is performed covertly, the system can be negatively affected for a long period of time. This paper develops a simulation stochastic cellular automata model of traffic on a circular two-lane road based on the Sakai-Nishinari-Fukui-Schadschneider (S-NFS) rules. In the presented model, in addition to ordinary vehicles, there are covertly counteracting vehicles; their task is to reduce the quantity indicators (such as traffic flux) of the transport system using special rules of behavior. Three such rules are considered and compared: two lane-changing rules and one slow-down rule. It is shown that such counteracting vehicles can affect the traffic flow, mainly in the region of the maximum of the fundamental diagram, that is, at average values of the vehicle density. In free-flowing traffic or in a traffic jam, the influence of the counteracting vehicle is negligible regardless of its rules of behavior.

## Introduction

The implementation of new transport systems within the scope of the smart city concept additionally carries an increasing threat of interference in various signal transmission processes with the aim of threatening the normal functioning of the transport systems [1–3]. The development of self-driving cars, in addition to the obvious advantages, also carries the risks of outside interference in the process of driving a vehicle. For example, a captured vehicle can be slowed down and stopped, causing a traffic jam to occur. Obviously, such a vehicle will be removed rather quickly by a tow truck and will not be able to affect traffic for a long time. Therefore, as an alternative, the captured vehicle can be controlled in such a manner as to stay in the transport system of the smart city for as long as possible, affecting the quantitative characteristics of transport flows. We will call such vehicles "counteracting vehicles".

Simulations are commonly used to address such problems—traffic simulation models are the main tools for analyzing transport systems. There are traffic simulation models of the macro-, meso- and micro-levels [4–6]. Quite significant progress in modeling traffic flows was achieved after the introduction of micro-level traffic simulation models based on cellular automata (CA) [7–9]. The advantages of this approach are the simplicity of implementation, since the model is completely discrete in both space and time, and computational efficiency, due to the natural parallelization of computations. The simplest CA model is based on the one-dimensional Wolfram rule 184 [9, 10]. Despite its elementary nature, this model allows one to simulate a rather complex topology of the road network, including entire cities, such as Geneva, for example [11]. Serious improvements to the CA model, such as a discrete set of velocities and taking the stochasticity into account (in case there is random braking), were introduced by Nagel and Schreckenberg [7, 8]. Although the classical models of Wolfram and Nagel-Schreckenberg did not correctly explain the phase transitions in Kerner's three-phase theory [12], modern CA models allow this to be done [13]. As a rule, all CA-based traffic simulation models described in the literature differ only in the stochastic component [14–17]. A disadvantage of CA-based models is the necessity to run the simulation multiple times to obtain statistically significant results.

The objective of this paper is to develop a traffic simulation model that can estimate the negative impact of counteracting vehicles on the overall flux of a transport system. We will consider three potential rules of behavior for counteracting vehicles: two lane-changing rules and one slow-down rule.

## Simulation model

We will consider a model of a closed circular two-lane road segment, which is a cellular automaton of (2, $N$) cells, with periodic boundary conditions (Fig. 1). The usual size of the cell along the movement direction corresponds to approximately 7.5 m. All vehicles in both lanes move in the same direction, as shown by the arrow, and can change their position between the lanes. The lane-changing rules are symmetrical to both lanes.

---

[1] A technical report

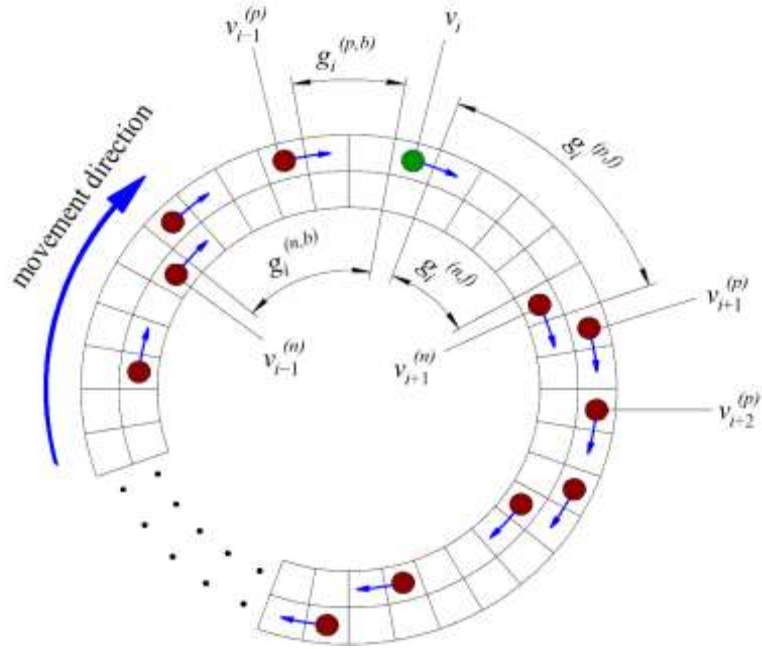

**Fig. 1.** Topology of the two-lane circular road transport network.

There can be two types of agents in the system: ordinary vehicles that move according to some usual driving rules, and counteracting vehicles, which have rules of behavior that are chosen to have a negative effect on the integral traffic indicators, such as flux. Counteracting vehicle behavior is individual, which means that each such vehicle has no information about whether another vehicle is ordinary or counteracting.

The time evolution of the transport system from step $t$ to step $t+1$ occurs in the following sequence. First, all vehicles determine the possibility of making a lane change, which occurs with a probability $P_{CL}$. Ordinary vehicles implement the lane-changing rules proposed in Ref. [18], which consist of two criteria:
  - incentive criterion:

$$g_i^{(n,f)} + v_{i+1}^{(n)} > v_i^{(p)} > g_i^{(p,f)} + v_{i+1}^{(p)}, \tag{1}$$

  - safety criterion:

$$v_i^{(p)} > v_{i-1}^{(n)} - g_i^{(n,b)}. \tag{2}$$

Here, $g_i^{(p,f)}$ is the distance (number of unoccupied sites) to the next vehicle in front of the current vehicle in the current lane; $g_i^{(n,f)}$ and $g_i^{(n,f)}$ are the distances to the vehicles in front of and behind the current vehicle, respectively, in the adjacent lane; $v_i^{(p)}$ is the velocity of the focal vehicle; $v_{i+1}^{(p)}$ is the velocity of the next vehicle along the current lane; $v_{i+1}^{(n)}$ and $v_{i-1}^{(n)}$ are the velocities of the vehicles in front of and behind the current vehicle, respectively, in the adjacent lane (see Fig. 1 for more details).

After the lane-changing step, all vehicles move along the selected lane until the end of the current time step. The motion of ordinary vehicles in current lane obeys the modified S-NFS model rules proposed by Nishinari, Fukui and Schadschneider [19] and modified later by Sakai [13], which are as follows. Let $v_i^{(0)}$ be the velocity of the vehicle $i$ at the beginning of the time step, and let $g_i$ be the distance to the next vehicle in the movement direction along the current lane (equal to $g_i^{(p,f)}$ in the previously described notation, see (1) and Fig. 1). The following rules are applied to each vehicle in parallel:
  1) Acceleration: if $g_i > G$ or $v_i^{(0)} \leq v_{i+1}^{(0)}$, then

$$v_i^{(1)} = \min[V_{\max}, v_i^{(0)} + 1]. \tag{3}$$

Here, $v_{i+1}^{(0)}$ is the velocity of the next vehicle along the current lane.
  2) Slow-to-start

$$V_i^{(2)} = \min[v_i^{(1)}, x_{i+s_i}(t-1) - x_i(t-1) - s_i], \tag{4}$$

with probability $q$, where $s_i = S$ with probability $r$ and $s_i = 1$ with probability $1 - r$.
  3) Quick start

$$V_i^{(3)} = \min[v_i^{(2)}, x_{i+s_i}(t) - x_i(t) - s_i]. \tag{5}$$

4) Random brake

$$V_i^{(4)} = \max[1, v_i^{(3)} - 1], \qquad (6)$$

with probability $1 - p_i$, where $p_i$ is defined as follows: if $g_i > G$, then $p_i = P_1$. Otherwise, $p_i = P_2$ if $v_i^{(0)} < v_{i+1}^{(0)}$, $p_i = P_3$ if $v_i^{(0)} = v_{i+1}^{(0)}$ and $p_i = P_4$ if $v_i^{(0)} > v_{i+1}^{(0)}$.

5) Avoid collision

$$V_i^{(5)} = \min[v_i^{(4)}, x_{i+1}(t) - x_i(t) - 1 + v_i^{(4)}]. \qquad (7)$$

6) Moving forward

$$X_i(t+1) = x_i(t) + v_i^{(5)}. \qquad (8)$$

Here, $x_i(t)$ is the position of vehicle $i$ at time $t$, and $g_i = x_{i+1}(t) - x_i(t)$. The velocity at the beginning of the current time step $v_i^{(0)}$ is equal to the velocity $v_i^{(5)}$ from the previous step:

$$v_i^{(0)}(t) = v_i^{(5)}(t-1) = x_i(t) - x_i(t-1).$$

The quantities $P_{CL}$, $G$, $q$, $r$, $S$, $P_1$, $P_2$, $P_3$ and $P_4$ are model parameters.

The counteracting vehicle must comply with the same safety and moving rules as ordinary vehicles (e.g., (2), (7) and (8)) but may act differently in other situations. The following three rules are considered below.

*First lane-changing rule.* The difference in the behavior of the counteracting vehicle is that it uses a different lane-changing incentive criterion (instead of (1)):

$$v_i^{(p)} < v_{i-1}^{(n)} \text{ and } v_i^{(p)} < g_i^{(n,f)} + v_{i+1}^{(n)}. \qquad (9)$$

If a vehicle $i - 1$, driving in an adjacent lane behind the counteracting vehicle $i$, is moving faster than the counteracting vehicle $i$, vehicle $i$ changes to this lane in front of vehicle $i - 1$.

*The second lane-changing rule.* For this rule, the incentive criterion is formulated as follows:

$$V_{i-1}^{(p)} < v_{i-1}^{(n)} \text{ and } v_i^{(p)} < g_i^{(n,f)} + v_{i+1}^{(n)}. \qquad (10)$$

In this case, the velocity of the following vehicle in the current lane is compared to the velocity of the following vehicle in the adjacent lane.

For both of these rules, the counteracting vehicle always change lanes when there is such a possibility, which means that $P_{CL} = 1$ for counteracting vehicles.

*Slow-down rule.* As mentioned before, the counteracting vehicle cannot slow down noticeably, as this will quickly revel its intentions. Hence, one of the possible slow-down rules can be formulated in the following way:

$$\text{If } (v_i^{(4)} = v_{i+1}^{(4)} \text{ and } v_i^{(4)} > v_{\min} \text{ and } g_i < G) \text{ then } v_i^{(4)} = v_i^{(4)} - 1. \qquad (11)$$

This means that if the counteracting vehicle is driving fast enough and has the same velocity as a close preceding vehicle ("close" means that distance to the preceding vehicle is acceptable for the application of the random brake rule), then it slows down for one discrete velocity step. This rule is applied after steps (3)–(6) are performed.

## Results and discussion

In the numerical experiments, the following values of the model parameters were assumed: $P_{CL} = 0.5$, $G = 15$, $S = 2$, $q = 0.99$, $r = 0.99$, $P_1 = 0.999$, $P_2 = 0.99$, $P_3 = 0.98$, $P_4 = 0.01$. The maximum vehicle velocity is $V_{\max} = 5$. The minimum velocity for applying the slow-down rule for counteracting vehicles is $v_{\min} = 3$. The number of lanes was set to $n_L = 2$ and the number of cells in each lane to $K_L = 10^3$. Each episode (single simulation) was divided into two periods: a run-up period, to attain the steady state of the system, and an observation period, in which simulation results are evaluated. The number of time steps of the run-up period is $T_{rup} = 4500$ and the number of time steps of the observation period is $T_{op} = 2500$.

The fraction of counteracting vehicles $R$ (a ratio of the counteracting vehicles to the total number of vehicles in the system) was varied in the range from 0 to 0.6 with a step of 0.3. For example, $R = 0.6$ corresponds to the situation where 60% of all vehicles in the system are counteracting.

Fig. 2 shows the fundamental diagrams (normalized flux $q$ versus normalized vehicle density $\rho$) and lane change frequencies $r$ versus the normalized vehicle density $\rho$ for three cases: $R = 0$ (no counteracting vehicles), $R = 0.3$ and $R = 0.6$. Each point in the graphs shown is the result of a single episode of the simulation (simulating the system evolution for $T_{rup} + T_{op}$ time steps). Fig. 3a and Fig. 3b correspond to the first lane-changing rule (9), Fig. 3c and Fig.3d correspond to the second lane-changing rule (10) and Fig. 3e and Fig. 3f correspond to the slow-down rule (11) for counteracting vehicles.

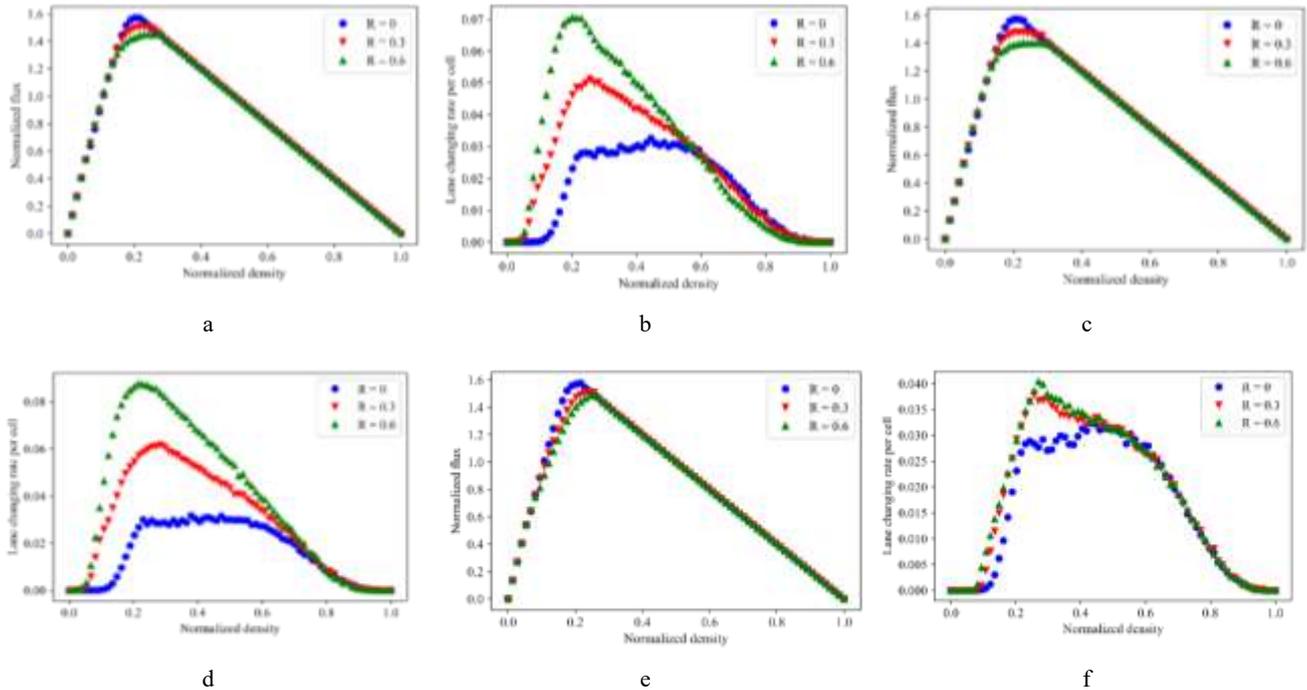

**Fig. 2.** Fundamental diagrams (a, c, and e) and frequencies of lane change (b, d and f) for different fractions of counteracting vehicles *R* in the situations involving the first lane-changing rule (a and b), second lane-changing rule (c and d) and slow-down rule (e and f) for counteracting vehicles.

All three fundamental diagrams (Fig. 3a, Fig. 3c and Fig. 3e) show that in the regions of free flow (i.e., $\rho < 0.1$) and jammed flow (i.e., $\rho > 0.4$), the presence and behavior of counteracting vehicles does not affect the overall flux. This can be easily explained: in a free flow, each vehicle moves like there are no other vehicles, and in a jammed flow, there is not enough space for maneuvers. Hence, the only vehicle density range in which counteracting vehicles can affect the flux is about $0.1 < \rho < 0.4$.

The second lane-changing rule of counteracting vehicles has a slightly stronger impact on flux than the first lane-changing rule. It can be also mentioned that the effect of the slow-down rule is shifted to the left (closer to the free-flow area of the diagram) since the counteracting vehicles (which obviously move slower on average in this case) are also taken into account in the evaluation of the normalized flux.

The lane-changing frequency curves in Fig. 3b and Fig. 3d reveal more clearly that counteracting vehicles perform most of their lane changing operations in the middle density region, where $0.1 < \rho < 0.4$. Fig. 3f shows that in order to avoid the negative effects created by counteracting vehicles, all vehicles in the system start to change lanes slightly more frequently.

## Conclusions

It is shown that for a two-lane cyclic transport system with the possibility of lane changing, the presence of covertly counteracting vehicles affects the traffic flow for middle vehicle density values, i.e., around the transition from free traffic to dense traffic. In the above region, the presence of counteracting vehicles reduces the maximum flux of the system. Of the three possible rules of behavior for counteracting vehicles, the lane-changing rule in which one needs to compare the velocities of two following vehicles has the most impact on the system flux. At the same time, covert slowing down has much smaller impact on the system flux.